\documentstyle[preprint,aps]{revtex}
\preprint{IMSc/97/12/41}
\begin{document}
\draft 
\title{Three flavor implications of CHOOZ result}
\author
{Mohan Narayan, G. Rajasekaran }
\address
{Institute of Mathematical Sciences, Chennai 600 113, India.}
\author
{S. Uma Sankar}  
\address
{Department of Physics, I.I.T. , Powai, Mumbai 400076, India}
\date{\today}
\maketitle

\begin{abstract}
We analyze the recent result of the CHOOZ collaboration in the context of
mixing and oscillations between all the three neutrino flavors. 
If one assumes
the  hierarchy among the vacuum 
mass eigenvalues  
$\delta_{21} \ll \delta_{31}$  where $\delta_{21}= \mu_2^2-\mu_1^2$ and
$\delta_{31}= \mu_3^2-\mu_1^2$,  
then  the CHOOZ result puts a
strong constraint on the allowed values of the (13) mixing angle $\phi$.
It is also shown 
that in light of the CHOOZ result, the maximum contribution of the  
$\nu_{\mu} \leftrightarrow \nu_{e}$ oscillation 
channel to the atmospheric neutrino anamoly is less than 9 percent, 
thus demonstrating that the atmospheric neutrino anamoly is mainly due to
$\nu_{\mu} \leftrightarrow \nu_{\tau}$ oscillations.
Most 
importantly the CHOOZ result now excludes a large part of the three flavor
parameter space which was previously allowed as solutions to the 
solar and atmospheric
neutrino problems.
\end{abstract}
\vspace{0.5cm}
\narrowtext

The CHOOZ collaboration, which searches for signals of
$\bar{\nu}_{e} \rightarrow  \bar{\nu}_{x}$ oscillations,
where $x$ can be any other flavor, in the disappearance 
mode of the original flavor
has recently reported the results of its 
first run \cite{chc}. They see no
evidence of oscillations of the original flavor.
They have analyzed their results assuming two flavor
oscillations between $\nu_e$ and another flavor and 
gave an exclusion plot in the parameter space spanned by 
the mass squared difference $\Delta m^2$ and the  
mixing angle $\theta$. 
Their main result
is that for $\Delta m^2 > 3 \times 10^{-3} eV^2$, 
$\sin^2 (2 \theta)$ must be
less than 0.18. While this is a strong constraint, 
we remark that it has to
be confirmed by an independent experiment. 
Nevertheless we may ask what are
the consequences if we accept the CHOOZ result. 
   
We reinterpret the CHOOZ result in terms of oscillations
between the three active neutrino flavors. This is a more
realistic framework because it is established that there 
are three light neutrino flavors whose interactions are
prescribed the Standard Model. It is more natural to assume
that all three of the light neutrinos mix with one another.
 
The flavor eigenstates are related to the 
mass eigenstates by 
\begin{equation}
\left[ \begin{array}{c} 
\nu_e \\ \nu_\mu \\ \nu_\tau \end{array} \right] 
= U 
\left[ \begin{array}{c} 
\nu_1 \\ \nu_2 \\ \nu_3 \end{array} \right]. 
\end{equation}
Here we can take, without loss of generality, that
$m_3 > m_2 > m_1$.
The unitary matrix $U$ can be parametrized as
\begin{equation}
U = U^{23} (\psi) \times U^{phase} \times U^{13} (\phi)
\times U^{12} (\omega),
\label{eq:defUv}
\end{equation}
where $U^{ij} (\theta_{ij})$ is the two flavor mixing matrix between 
the ith and jth mass eignestates with the mixing angle $\theta_{ij}$.
For simplicity, we neglect the CP violation and set 
$U^{phase} = I$.
The vacuum oscillation probability for a neutrino of flavor
$\alpha$ to oscillate into a neutrino of
flavor $\beta$ is given by
\begin{eqnarray}
P^0_{\alpha \beta} & = & 
\left( U_{\alpha 1} U_{\beta 1} \right)^2 + 
\left( U_{\alpha 2} U_{\beta 2} \right)^2 + 
\left( U_{\alpha 3} U_{\beta 3} \right)^2 +  
2 \ U_{\alpha 1} U_{\alpha 2} U_{\beta 1} U_{\beta 2}
    \cos \left( 2.53 \frac{d \ \delta_{21}}{E} \right) + \nonumber \\
& & 2 \ U_{\alpha 1} U_{\alpha 3} U_{\beta 1} U_{\beta 3}
    \cos \left( 2.53 \frac{d \ \delta_{31}}{E} \right) + 
2 \ U_{\alpha 2} U_{\alpha 3} U_{\beta 2} U_{\beta 3}
    \cos \left( 2.53 \frac{d \ \delta_{32}}{E} \right),  
\label{eq:Palbe}   
\end{eqnarray}
where $d$ is the distance travelled in meters, 
$E$ is in MeV, and mass squared
differences are in eV$^2$. We may also note the vacuum oscillation
probabilities are same as in eq.~(\ref{eq:Palbe})
for the case of antineutrinos because CP violation is neglected. 
If we assume the hierarchy among the neutrino mass eigenstates
$\delta_{31} \gg \delta_{21}$, 
and that $\delta_{21}$ is about $10^{-5} eV^2$, which  
is required to fit solar neutrino data \cite{halg}, then the 
oscillatory term involving
$\delta_{21}$ can be set to one.
The oscillation probability relevant for the CHOOZ
experiment is the electron neutrino  survival probability 
$P_{e e}$ which is easily computed from eq.~(\ref{eq:Palbe}) to be 
\begin{equation}
P_{e e} =
1 - \sin^2 2 \phi  \sin^2 \left(1.27 \frac{d \ \delta_{31}}{E}
\right). 
\end{equation} 
Notice the interesting point that this involves 
only the (13) mixing angle $\phi$, and
because of the heirarchy the (12) mixing angle 
$\omega$ disappears from the
probability.
So we reinterpret the CHOOZ result \cite{chc}, to be that 
for $\delta_{31} > 3 \times 10^{-3}$, 
 $\sin^2(2 \phi)$ must be less than 0.18, i.e
$\phi < 12.5^o$. 

We now estimate the maximum contribution of the 
$e - \mu$ channel to the 
atmospheric neutrino anomaly.
Since the relevent $\delta_{31}$ is about 
$10^{-2} eV^2$, matter effects are
negligible for the problem \cite{nru}. 
Hence the relevant probability is
the vacuum $\nu_{e} \leftrightarrow \nu_{\mu}$ oscillation probability,
\begin{equation}
P_{\bar{\mu} \bar{e}} = P_{\mu e} =
\sin^2 2 \phi \sin^2 \psi \sin^2 \left(1.27 \frac{d \ \delta_{31}}{E}
\right). 
\end{equation} 
Note that both $\phi$ and $\psi$ have to be non-zero for $P_{\mu e}$ to
be non-zero, and also the oscillation length corresponding to $\delta_{21}$
does not contribute to the atmospheric neutrino problem \cite{nru}.
Now solutions to Kamiokande atmospheric 
neutrino data \cite{nru,nmru} require 
a value of $\psi \geq 45^o$. The average contribution of the
oscillatory term is 0.5. Therefore using the CHOOZ result
that the maximum value of $\sin^2(2 \phi)$ allowed is 0.18 
we get
\begin{equation}
P^{max}_{\mu e} \leq 1.0 \times 0.18 \times 0.5  = 0.09  
\end{equation}
which is less than 9 percent. 
Hence the atmospheric neutrino anomaly is driven
almost completely by $\nu_{\mu} \leftrightarrow \nu_{\tau}$ oscillations.
The $\nu_{e} \leftrightarrow \nu_{\tau}$ 
conversion probability is given by
\begin{equation}
P_{\bar{e} \bar{\tau}} = P_{e \tau} =
\sin^2 2 \phi \cos^2 \psi \sin^2 \left(1.27 \frac{d \ \delta_{31}}{E}
\right) \label{eq:lsndosc}.
\end{equation} 
Since $\psi > 45^o$, we find that the $e - \tau$ conversion 
probability is less than 5 percent,
i.e. the electron neutrino flux is hardly
converted to other flavors, which is what is experimentally observed.

Lastly we incorporate the CHOOZ constraints 
on our previous fits to solar and
atmospheric neutrino data, and so we reproduce 
the plots from our earlier works, with
the  constraints coming from the CHOOZ results shown on them.
In Fig.1, the light contours enclose the parameter region in $\phi-\psi$
plane allowed by the binned multi-GeV data of Kamiokande with 
$1.6 \ \sigma$ error bars. The 
present CHOOZ constraint has been shown as a thick vertical line,
with the region to the right of it being excluded.
Fig.2 shows the allowed region in the $\phi-\delta_{31}$ 
plane from the same analysis, with the CHOOZ constraint again
being shown as a thick vertical line \cite{nru}. 
Fig.3 and Fig.4 show
the previously allowed regions by the solar neutrino data
in $\phi-\omega$ and 
$\phi-\delta_{21}$ planes respectively 
along with the new constraint \cite{nmru}.

Note the fact that $\phi$ being the angle 
which connects the solar neutrino
parameter space spanned by $\omega, 
\phi$, and $\delta_{21}$ with the atmospheric
neutrino space spanned by $\phi, \psi$, and 
$\delta_{31}$, the constraint on $\phi$
also translates into a  strong  constraint 
in the solar neutrino parameter
space. Now observe what is probably the most 
important consequence of the CHOOZ
result. The fact that $\phi$, the link 
between the solar and the atmospheric
neutrino problems is constrained to be small 
implies that {\it the solar neutrino
problem can be essentially viewed as a two flavor 
$\nu_{e} \leftrightarrow \nu_{\mu}$
oscillation phenomena, 
and the atmospheric neutrino problem essentially as a two
flavor $\nu_{\mu} \leftrightarrow \nu_{\tau}$ 
oscillation phenomena even in a three
flavor framework.}

In conclusion the recent CHOOZ result limits 
the $\nu_{\mu}\leftrightarrow \nu_{e}$
contribution to the atmospheric 
neutrino anomaly as a function of the
(13) mixing angle $\phi$,  
establishes the fact that the atmospheric
neutrino anomaly is mainly 
$\nu_{\mu} \leftrightarrow \nu_{\tau}$ i.e vacuum
oscillations, and excludes large parts of 
the parameter space previously
allowed as solutions to
solar and atmospheric neutrino data.

We thank M.V.N. Murthy and Rahul Sinha for discussions.

\begin{figure}
\caption{Allowed parameter region in $\phi-\psi$ plane by   
Kamiokande binned multi-GeV data with $1.6 \ \sigma$ error
bars (light lines) and the new constraint by CHOOZ (thick line). 
}\label{Fig. 1}
\end{figure}

\begin{figure}
\caption{Allowed parameter region in $\phi-\delta_{31}$ plane by   
Kamiokande binned multi-GeV data with $1.6 \ \sigma$ error
bars (light lines) and the new constraint by CHOOZ (thick line). 
}\label{Fig. 2}
\end{figure}

\begin{figure}
\caption{Allowed parameter region in $\phi-\omega$ plane by   
solar neutrino data with $1.6 \ \sigma$ error
bars (crosses) and the new constraint by CHOOZ (thick line). 
}\label{Fig. 3}
\end{figure}

\begin{figure}
\caption{Allowed parameter region in $\phi-\delta_{21}$ plane by   
solar neutrino data with $1.6 \ \sigma$ error
bars (crosses) and the new constraint by CHOOZ (thick line). 
}\label{Fig. 4}
\end{figure}

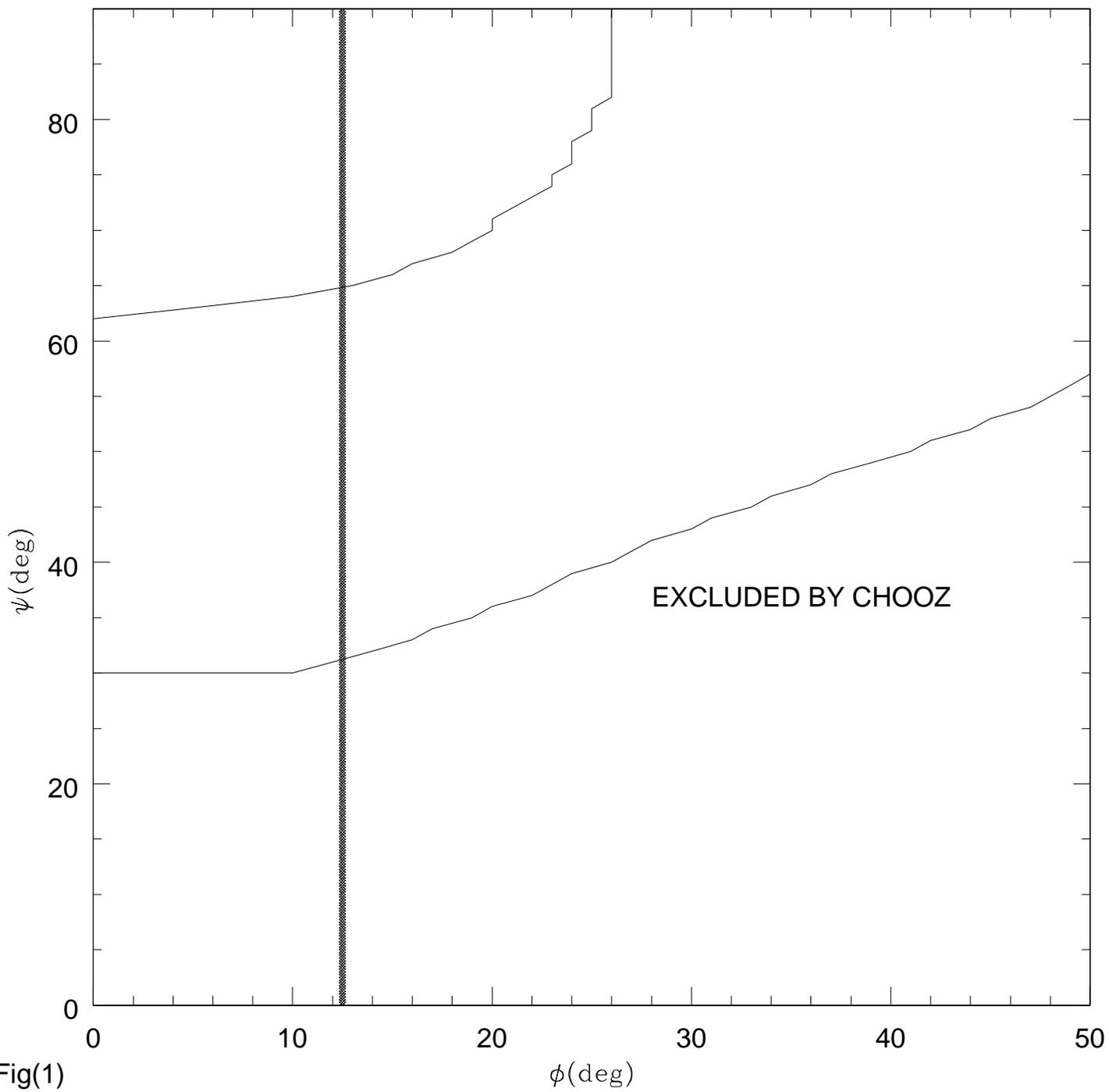
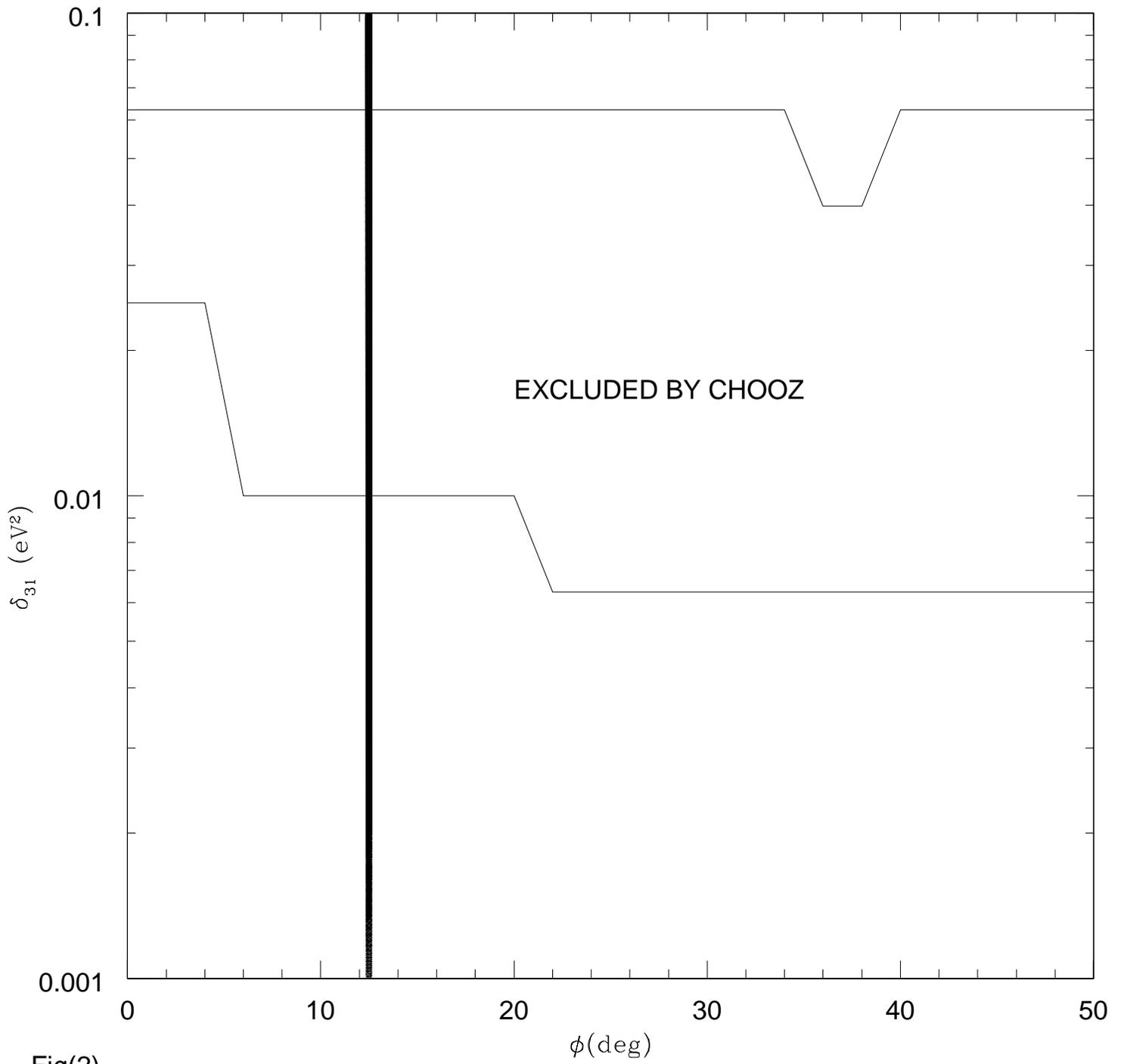
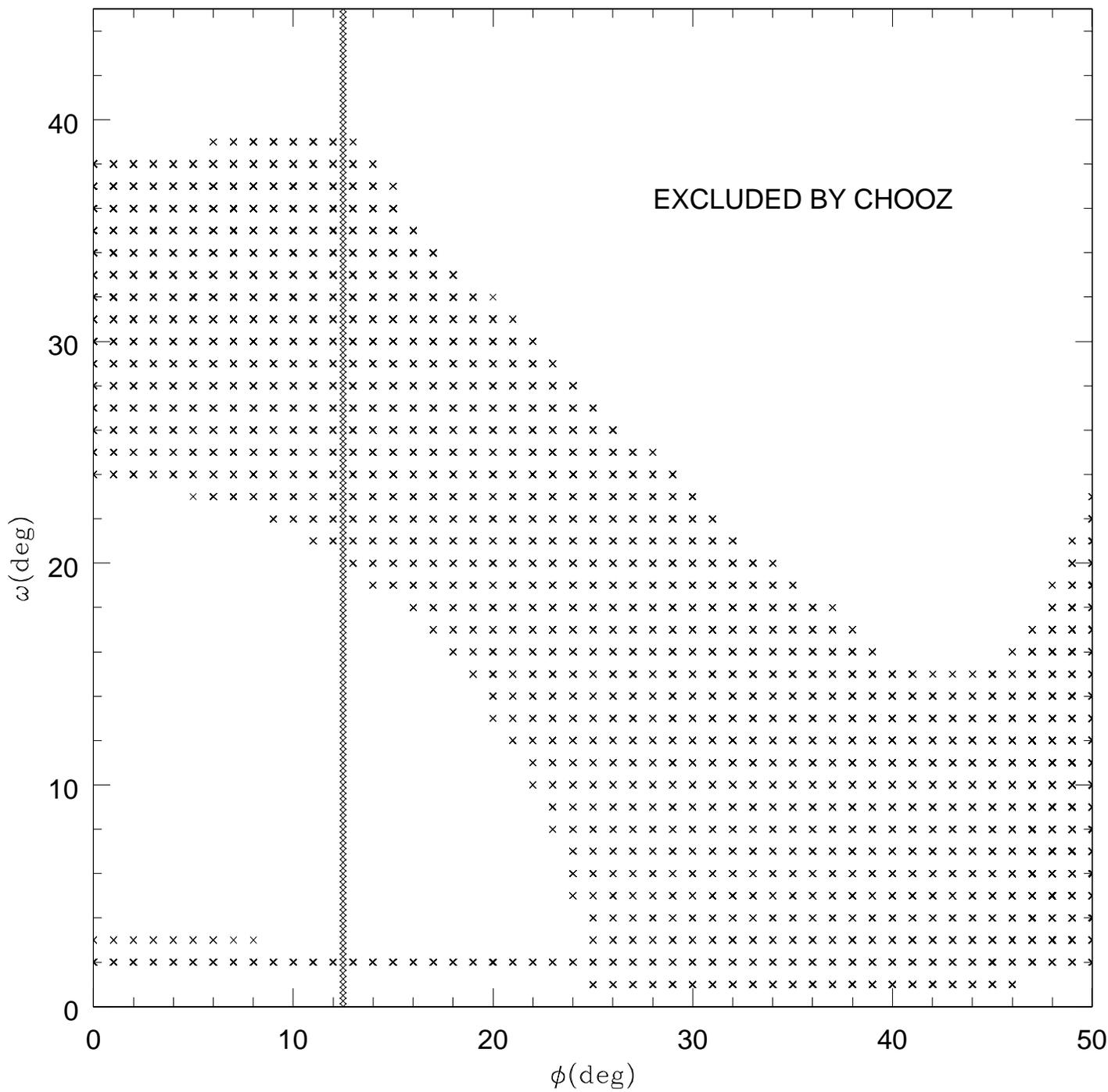
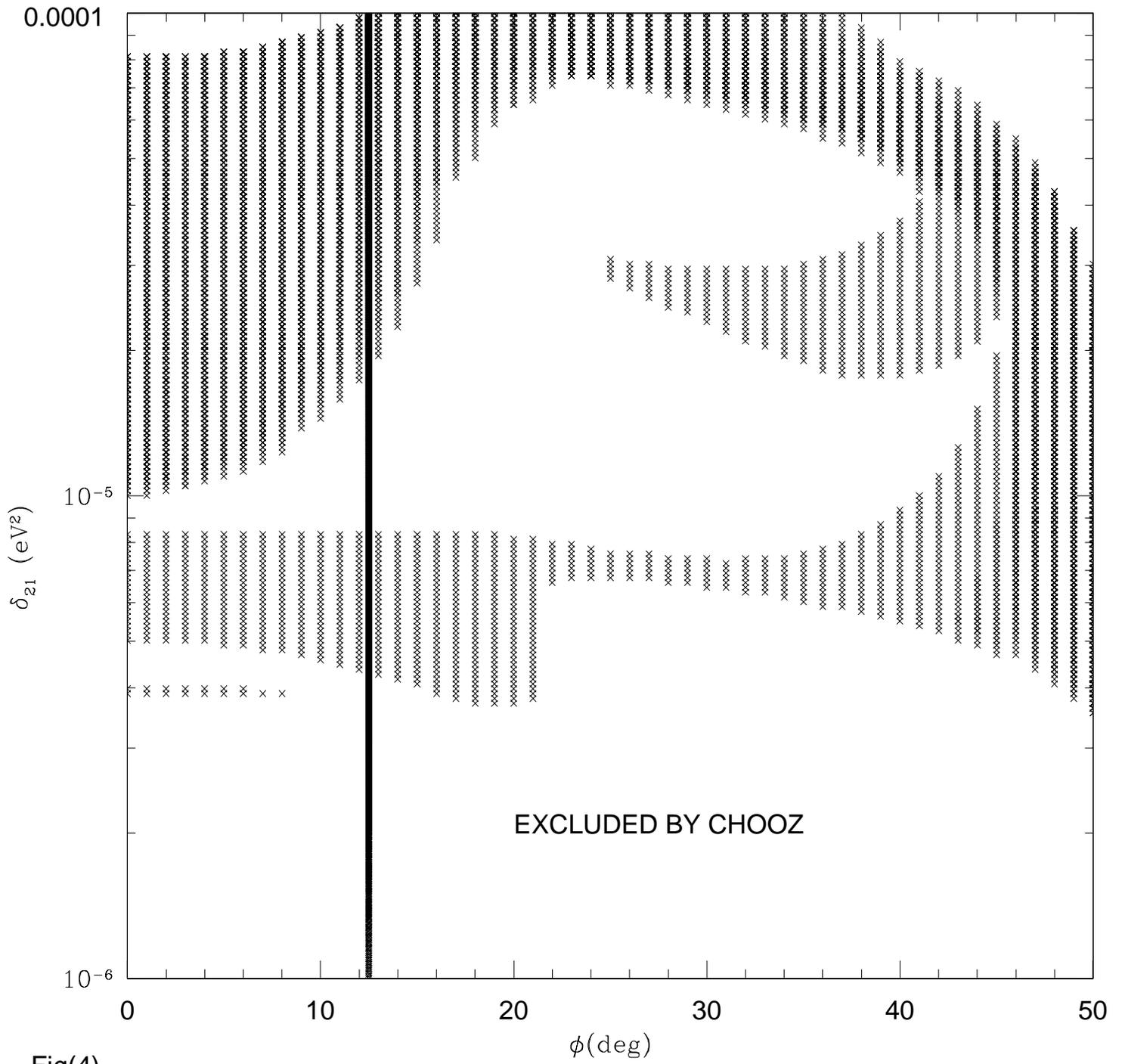
\end{document}